\documentclass[12pt]{amsart}
\usepackage{amssymb}

\usepackage[dvips]{graphicx}
\usepackage{epsfig}

\newcommand{\Ha}{{\mathcal H}}
\newcommand{\Ka}{{\mathcal K}}
\newcommand{\jed}{{\mathbf 1}}
\newcommand{\C}{{\mathbb C}}
\newcommand{\E}{{E}}
\newcommand{\gwia}{^{\star}}
\newtheorem{theo}{Theorem}

\begin{document}

\title[Limit distributions of many--particle spectra]%
{Limit distributions of many--particle spectra \\ and $q$-deformed
Gaussian variables}

\author{Piotr \'Sniady}

\address{Institute of
Mathematics, University of Wroc\l{}aw, pl.~Grunwaldzki 2/4, 50-384
Wroc\l{}aw, Poland \\ email: Piotr.Sniady@math.uni.wroc.pl}

\maketitle

\begin{abstract}
We find the limit distributions for a spectrum of a system of $n$ particles
governed by a $k$--body interaction. The hamiltonian of this system is
modelled by a Gaussian random matrix.

We show that the limit distribution is a $q$--deformed Gaussian distribution
with the deformation parameter $q$ depending on the value of the fraction
$\frac{k}{\sqrt{n}}$. The family of $q$--deformed Gaussian distributions
include the Gaussian distribution and the semicircular law; therefore our
result is a generalization of the results of Wigner \cite{W1,W2},
 Mon and French \cite{MF}.

\end{abstract}

\section{Introduction}
\subsection{The $k$--body interactions}
We say that the interaction of $n$ particles is a $k$--body interaction if
it can be treated as a sum of independent interactions, each taking place in
a group of $k$ particles (the groups of course need not to be disjoint). The
integer number $k$ is called a rank of the interaction.

From basic principles of physics we would expect that the fundamental
interactions should take place only in pairs of particles; by taking into
account higher order interactions between carriers of interaction we
see that also $k$--body interactions are possible if $k$ is a small integer.

Perhaps even better examples of a $k$--body interaction we obtain by
considering models of complex physical systems which were simplified by
disregarding some components. In such models effective hamiltonians contain
also $k$--body interactions ($k>2$) in order to preserve the effects
connected to the disregarded components.

Nevertheless we would expect
that $k$, the rank of the interaction should be relatively small.

\subsection{Random matrix models for complex quantum systems}
\subsubsection{Wigner's model}
The first and the simplest random matrix model for complex quantum systems
such as atomic nuclei was proposed by Wigner \cite{W1}. In this model a
hamiltonian of the system is represented by a large hermitian matrix
$(a_{ij})_{1\leq i,j\leq N}$ which entries are complex Gaussian random
variables. This model can be heuristically justified as follows: in an
sufficiently complex physical system the matrix elements of the hamiltonian
should be very complicated as well and therefore can be regarded as random.

In the Wigner's model the matrix element between any two physical
states can take nonzero values, what
for quantum systems consisting of $n$ parts means that the system is
governed by an $n$--body interaction, what in the light of the preceding
discussion is physically doubtful.

\subsubsection{Random matrix model for $k$--body interaction}
The following more realistic model was proposed \cite{BF1,BF2,FW1,FW2,WF} in
which a $k$--body interaction hamiltonian of $n$ particles is modeled by a
random matrix.

We consider a quantum system of $n$ distinguishable particles;
the Hilbert space $\Ha$ of the system is a tensor product of the Hilbert
spaces $\Ha_i$ ($1\leq i\leq n$) corresponding to particles. In fact as
``particles'' we can take also quantum statistical objects such as
collective excitations, holes, etc.

We assume that the rank of the effective interaction in our system is equal to
$k$ and therefore the hamiltonian $H$ of the system is a sum of hamiltonians
$H_A$ of the $k$--particle subsystems,
$$H=\sum_{A} H_A.$$
The sum is taken over all
sets $A\subset\{1,\dots,n\}$ which have exactly $k$ elements.

Due to the factorization of the Hilbert space $\Ha=\Ha_A\otimes\Ha_{A'}$
where $\Ha_A=\bigotimes_{i\in A} \Ha_i$, $\Ha_{A'}=\bigotimes_{i\not\in A}
\Ha_i$ we can write each hamiltonian $H_A$ as
$H_A=H_A^0\otimes\jed_{\Ha_{A'}}$ where $H_A^0:\Ha_A\rightarrow\Ha_A$ is a
selfadjoint operator and $\jed_{\Ha_{A'}}:\Ha_{A'}\rightarrow\Ha_{A'}$ is
the identity. Suppose that each of the particles has $s$ possible states and
therefore $\Ha_i=\C^s$ and $\Ha_A=\C^{s^k}$; we see that $H_A^0$ can be viewed as a
hermitian matrix with $s^k$ rows and columns.

Similarly as in the Wigner's model we shall assume that $H_A^0=(a^A_{i,j})$
is a hermitian random matrix, i.e. $a^A_{i,j}$ ($1\leq i\leq j\leq s^k$)
are complex Gaussian random variables with the mean $0$ and the covariance
$\E(a^A_{i,j} a^A_{k,l})=\E(a^A_{i,j} \overline{a^A_{l,k}})=
\frac{1}{{n \choose k} s^k} \delta_{il} \delta_{jk}$. We assume that
the entries of different hamiltonians $H_A$ are independent.

The above considered random matrices are related to Gaussian unitary
ensemble, but it is easy to write a version which is related to Gaussian
orthogonal or symplectic ensemble.

\subsection{Overview of the article}
The goal of this paper is to investigate the limit distribution of the above
spectra when $n$ tends to infity and the rank of the interaction $k(n)$
depends in a certain way on $n$.

A classical result of Wigner \cite{W1,W2} states that if $k(n)=n$ then the
distribution of the eigenvalues converges to the semicircular distribution
(see Fig.\ \ref{fig:wigner}). On the other hand Mon and French \cite{MF}
showed if $k$ is constant and $n$ converges to infinity then the
distribution of eigenvalues converges to the Gaussian distribution (see
Fig.\ \ref{fig:gauss}).

In this article we show that in the intermediate cases when $1\ll
k(n) \ll n$ the limit distributions are given by so--called $q$--deformed
Gaussian distributions.

%

\section{The $q$--deformed Gaussian random variables}
\label{sec:qdeform}
For overview articles on $q$--deformed commutation relations and
$q$--deformed Gaussian variables we refer to
\cite{FB,BS0,BS,BKS,vLM,Sn1,Sn2,Sp}. In this article we will consider only
one $q$--deformed Gaussian variable, what simplifies the discussion
significantly.

Let us consider a real number $-1<q<1$. One says that an operator $a$
acting on a some Hilbert space $\Ka$ and its adjoint $a\gwia$ fulfil the
$q$--deformed commutation relations if
$$a a\gwia-q a\gwia a=\jed,$$ where $\jed$ is the identity operator.

Suppose that there exists a unital vector $\Omega\in\Ka$ such that
$$a \Omega=0.$$
A vector with such a property is called a vacuum. In such a setup we can
introduced a (non--commutative) expectation value $\tau(X)=\langle \Omega,
X\Omega\rangle$.

By the $q$--deformed Gaussian random variable we mean the operator
$a+a\gwia$.

The $q$--deformed Gaussian distribution is a compactly supported
probabilistic measure
$\nu_q$ on the real line with a property that for each natural number $n$ we
have
$$\int x^n d\nu_q(x)=\tau [(a+a\gwia)^n]=\langle\Omega,
(a+a\gwia)^n\Omega\rangle. $$ It can be proven \cite{Sz} that for
$-1<q<1$ this measure is supported on the interval
$\left[-\frac{2}{\sqrt{1-q}},\frac{2}{\sqrt{1-q}} \right]$ and the
density of this measure is given by $$\nu_q(dx)=\frac{1}{\pi}
\sqrt{1-q} \sin \theta \prod_{n=1}^{\infty} (1-q^n) |1-q^n e^{2 i
\theta}|^2 dx, $$ where $\theta\in[0,\pi]$ is defined by
$x=\frac{2}{\sqrt{1-q}} \cos \theta $.

\begin{figure}  
\centering
\psfig{file=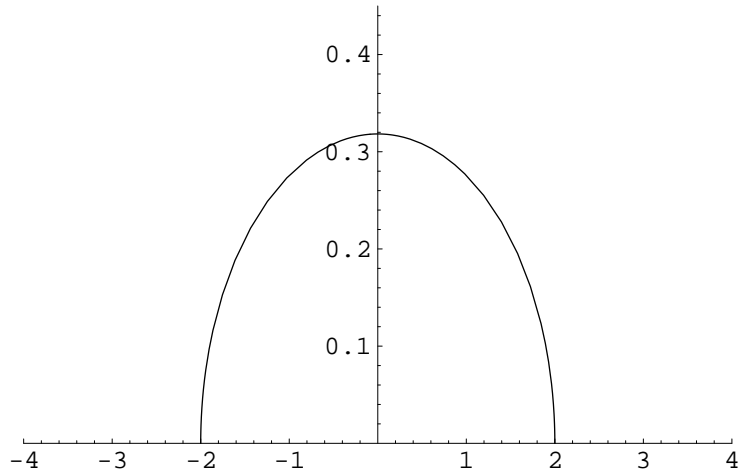,width=0.8 \textwidth}
\caption{Wigner semicircular law, corresponding to $q=0$.}
\label{fig:wigner}
\end{figure}

\begin{figure}  
\centering
\psfig{file=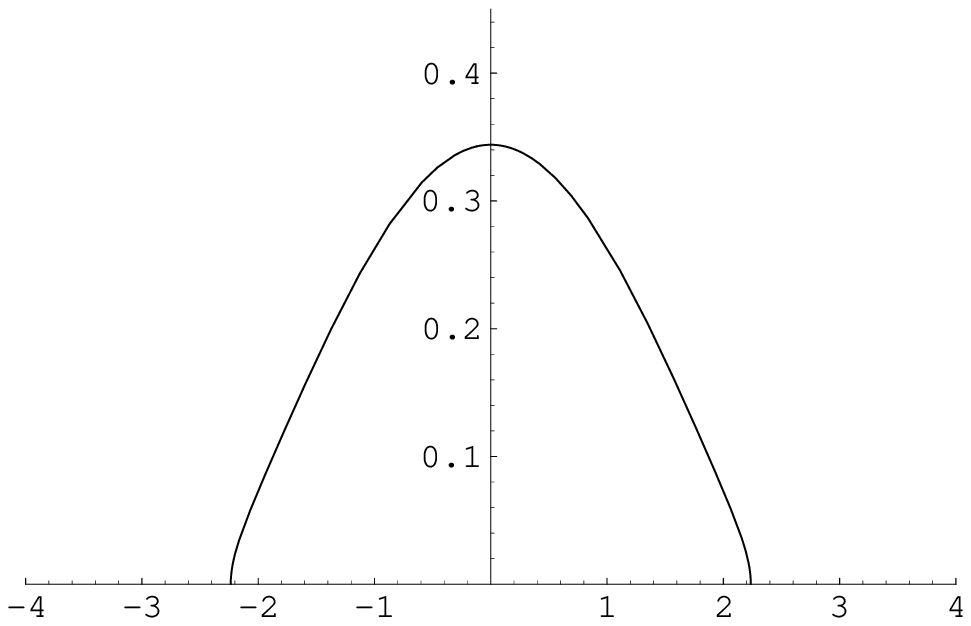,width=0.8 \textwidth}
\caption{The $q$--deformed distribution for $q=0.2$.}
\label{fig:krzywa02}
\end{figure}

\begin{figure}  
\centering
\psfig{file=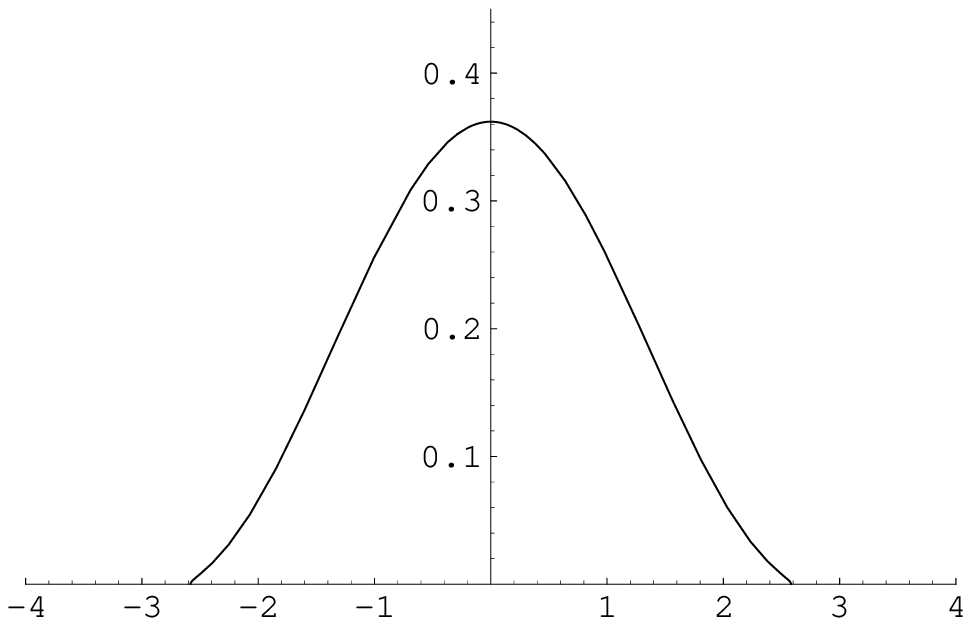,width=0.8 \textwidth}
\caption{The $q$--deformed distribution for $q=0.4$.}
\label{fig:krzywa04}
\end{figure}

\begin{figure}  
\centering
\psfig{file=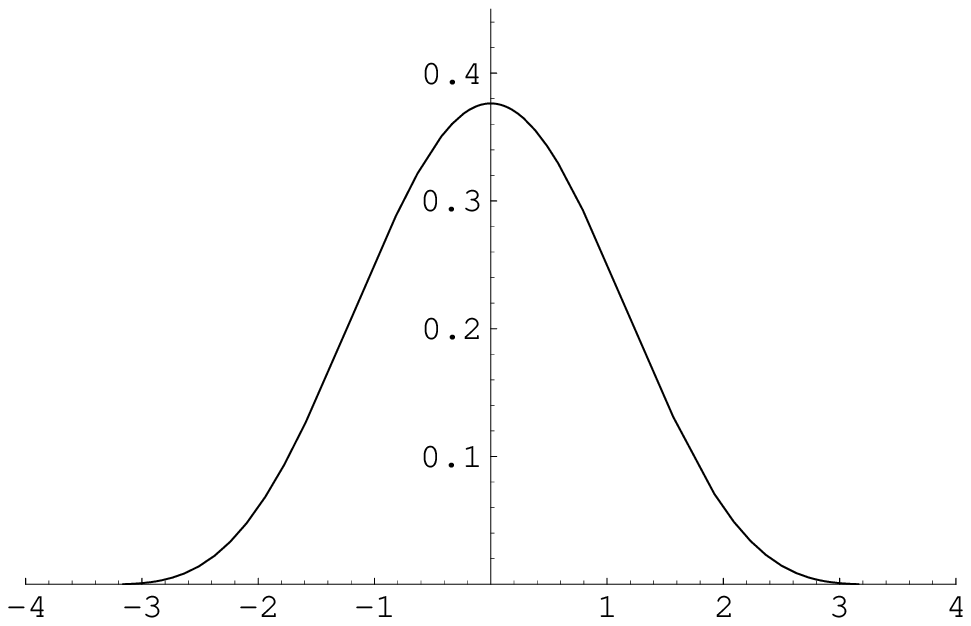,width=0.8 \textwidth}
\caption{The $q$--deformed distribution for $q=0.6$.}
\label{fig:krzywa06}
\end{figure}

\begin{figure}  
\centering
\psfig{file=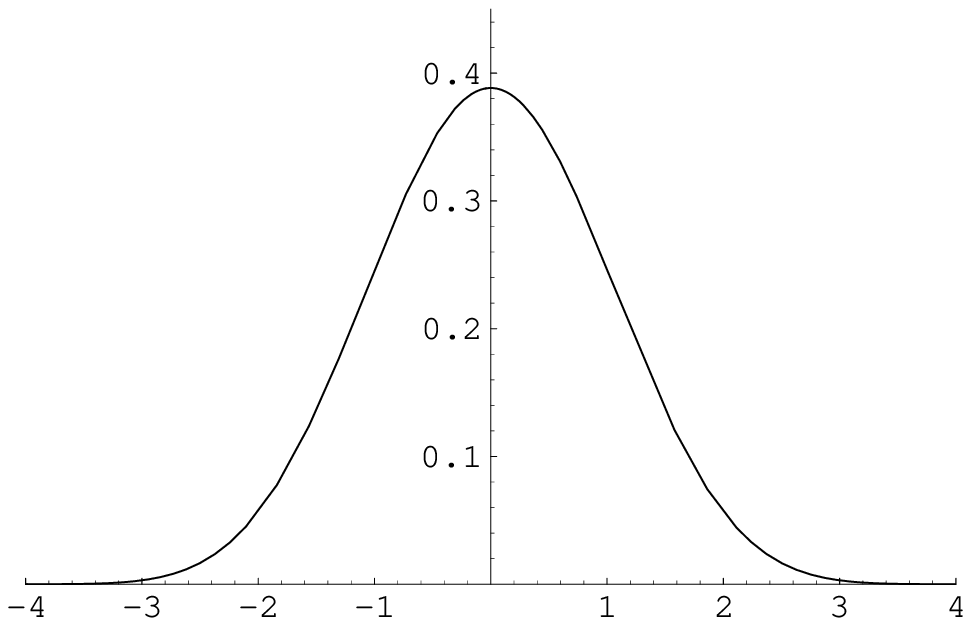,width=0.8 \textwidth}
\caption{The $q$--deformed distribution for $q=0.8$.}
\label{fig:krzywa08}
\end{figure}

\begin{figure}  
\centering
\psfig{file=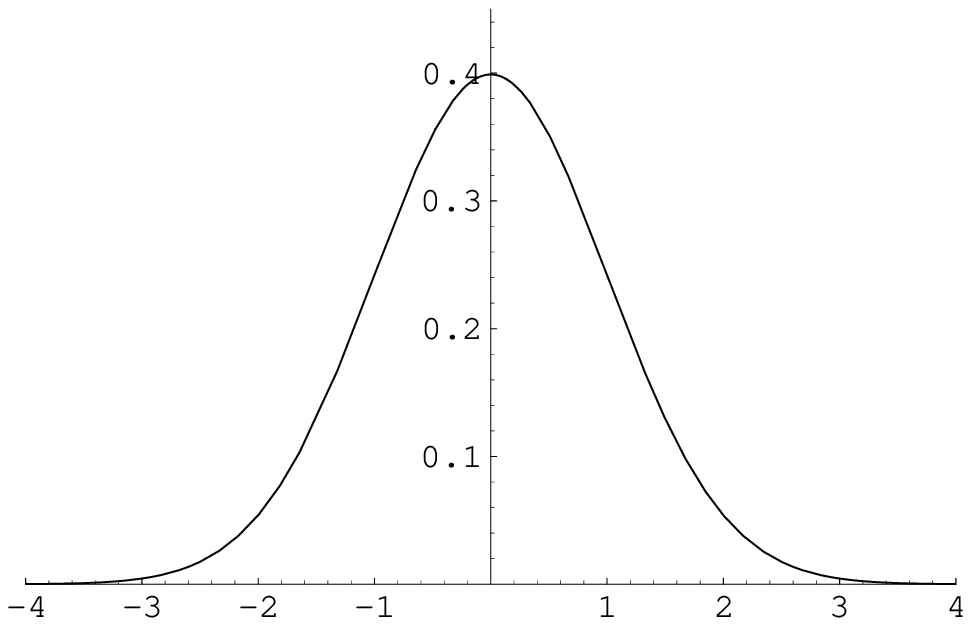,width=0.8 \textwidth}
\caption{The Gaussian distribution, corresponding to $q=1$.}
\label{fig:gauss}
\end{figure}

It turns out that for $q=1$ the $q$--deformed Gaussian distribution
coincides with the standard Gaussian distribution (see Fig.\ \ref{fig:gauss}).
For $q=0$ the $q$--deformed Gaussian distribution coincides with
the semicircular distribution of Wigner (see Fig.\ \ref{fig:wigner}).
The $q$--deformed Gaussian distributions for intermediate values of
the deformation paramater $q$ are presented on the Figures
\ref{fig:krzywa02}---\ref{fig:krzywa08}.

\section{The distribution of eigenvalues}
\label{sec:distrib}
The connection of the random matrix models considered in Introduction with
$q$--deformed Gaussian variables was found by the author in the following
theorem \cite{Sn2}.
\begin{theo}
If $\lim_{n\rightarrow\infty} \frac{k(n)}{\sqrt{n}}=c$, where $0\leq c\leq
\infty$ then the limit distribution of the eigenvalues of the hamiltonians
$H$ is the $q$--deformed Gaussian distribution with the parameter $q$ given
by $$q=\exp\left[-\left(1-\frac{1}{s^2} \right) c^2\right], $$
where $s$ is the number of one--particle states.
\end{theo}
It should not be a surprise to see that the above theorem contains results
of Wigner (in the model considered by him we have $k=n$ and
$\lim_{n\rightarrow \infty} \frac{k}{\sqrt{n}}=\infty$, $q=\exp[-\infty]=0$)
and of Mon and French ($k$ is constant, therefore $c=0$ and hence $q=1$).

We see that the square root from the number of particles is
the scale of the rank of the interaction in which the passage from the Gauss
law to the semicircular law occurs.

\section{Final remarks}
Particles considered in this article were distinguishable. The next
step would be to replace them by fermion or bosons. In the fermionic
case it would mean for example that $s$, the dimension of a
one--particle Hilbert space must be a function of $n$, the number of
particles. Simple arguments show that if $s(n)$ tends to infinity
faster than linearly, then the most of the states are not occupied
and Pauli exclusion principle does not affect our calculations. In
this case we can take $s=\infty$ and therefore
$$q=e^{-c^2}.$$

\section{Acknowledgements}
Research supported by State Committee for Scientific Research
(Komitet Bada\'n Naukowych) grant No.\ 2P03A00723; by EU Network
``QP-App\-li\-ca\-tions", contract HPRN-CT-2002-00729; by KBN-DAAD
project 36/2003/2004.

\end{document}